\documentclass[a4paper,11pt]{article}
\usepackage{pos}
\usepackage{subfig}
\usepackage{caption}
\usepackage{float}
\usepackage{cleveref}
\title{Development of an advanced SiPM camera for the Large Size Telescope of the Cherenkov Telescope Array Observatory}
 \ShortTitle{LST Advanced SiPM Camera}

\author*[a]{M. Heller}
\author[b]{T. Armstrong}
\author[c]{M.~Bellato}
\author[c]{A.~Bergnoli}
\author[c]{M.~Bernardos}
\author[g]{E.~Bernasconi}
\author[h]{A.~Biland}
\author[g]{E.~Charbon}
\author[c]{D.~Corti}
\author[a]{M.~Dalchenko}
\author[a]{D.~della Volpe}
\author[d,e]{D.~Depaoli}
\author[d]{F.~Di Pierro}
\author[a]{G.~Emery}
\author[i]{D.~Gascón}
\author[i]{S.~Gómez}
\author[c]{R.~López-Coto}
\author[c]{M.~Mariotti}
\author[a]{L.~D.~M.~Miranda}
\author[a]{T.~Montaruli}
\author[a]{A.~Nagai}
\author[c]{R.~Rando}
\author[j]{T.~Saito}
\author[f]{H.~Tajima}
\author[k]{K.~Zi{\c{e}}tara}

\affiliation[a]{DPNC - Universit\'e de Gen\`eve, 24 Quai Ernest Ansermet, CH-1211 Gen\`eve,  Switzerland}
\affiliation[b]{Aix Marseille Univ, CNRS/IN2P3, CPPM, Marseille, France}
\affiliation[c]{INFN Sezione di Padova, Via Marzolo, 8 - 35131 Padova, Italy}
\affiliation[d]{INFN Sezione di Torino, via P. Giuria, 1 - 10125 Torino, Italy}
\affiliation[e]{Università degli Studi di Torino, Dipartimento di Fisica, via P. Giuria, 1 - 10125 Torino, Italy}
\affiliation[f]{Institute for Space–Earth Environmental Research and Kobayashi–Maskawa Institute for the Origin of Particles and the Universe, Nagoya University, Nagoya 464-8601, Japan}
\affiliation[g]{AQUA, EPFL, Neuchâtel, Switzerland}
\affiliation[h]{ETH Zurich, Institute for Particle Physics, Otto-Stern-Weg 5, 8093 Zurich, Switzerland}
\affiliation[i]{Institute of Cosmos Sciencies - University of Barcelona, Mart\'i i Franqu\`es, 1, 08028, Barcelona, Spain}
\affiliation[j]{Institute for Cosmic Ray Research, The University of Tokyo, Japan}
\affiliation[k]{Astronomical Observatory, Jagiellonian University, ul. Orla 171, 30-244, Krak{\'o}w, Poland}

\forProj{CTA LST} 

\emailAdd{matthieu.heller@unige.ch}

\abstract{
Silicon photomultipliers (SiPMs) have become the baseline choice for cameras of the small-sized telescopes (SSTs) of the Cherenkov Telescope Array (CTA).
On the other hand, SiPMs are relatively new to the field and covering large surfaces and operating at high data rates still are challenges to outperform photomultipliers (PMTs). 
The higher sensitivity in the near infra-red and longer signals compared to PMTs result in higher night sky background rate for SiPMs. However, the robustness of the SiPMs represents a unique opportunity to ensure long-term operation with low maintenance and better duty cycle than PMTs.
The proposed camera for large size telescopes will feature $0.05^{\circ}$ pixels, low power and fast front-end electronics and a fully digital readout.
In this work, we present the status of dedicated simulations and data analysis for the performance estimation. The design features and the different strategies identified, so far, to tackle the demanding requirements and the improved performance are described.}

\FullConference{37$^{\rm{th}}$ International Cosmic Ray Conference (ICRC 2021)\\
		July 12th -- 23rd, 2021\\
		Online -- Berlin, Germany}

\begin{document}
\maketitle

\section{Introduction}

The first Large Size Telescope~\cite{lst_mazin}, LST-1,  installed at La Palma Observatory Roque de Los Muchachos, was inaugurated in Oct. 2018. It is a hundred meters away from the two MAGIC telescopes and close to the FACT small telescope, the first small telescope employing SiPMs.
The LST-1 is being commissioned and has been regularly taking data on gamma sources successfully since Nov. 2019. 
The construction of three additional LSTs at the La Palma site is ongoing. Once completed, the four LSTs will constitute, with five medium-sized telescopes (MSTs) at the CTA-North observatory.

To achieve the desired sensitivity and full-sky coverage, one array in the Northern and another in the Southern hemisphere including about 100 telescopes are planned for CTA. 
The Southern array has the best view of the Galactic Plane, which contains sources whose emission is most intense at energies below 1~TeV and in some cases can extend to the PeV region. 
Gammas with these energies can still reach us from distances of about  over 10~kpc (from the Galaxy), but would be absorbed for larger distances. 
It was considered to equip the Southern site with three types of telescopes with different mirror sizes: LSTs with mirrors of 23~m diameter, MSTs with mirrors of 12~m diameter, and small-sized telescopes (SSTs) with mirrors of 4~m diameter. Overall, the variable mirror size allows to cover the energy range from 20~GeV to $\sim 300$~TeV. 

A limiting factor of the current technology of gamma-ray telescopes concerns the duty cycle, which is limited to about 10\%, being  operation only possible during moonless nights without clouds, rain, sand bursts and with winds below $\sim 50$~km/h.

\section{Project goals and challenges}

The project aims at designing and building a new prototype camera for the LST in the 4 to 5 coming years. The ambition of the proposed design extends beyond the simple replacement of PMTs by SiPMs. While this transition would naturally provide higher duty cycle and robustness, very careful design is required to reach a performance gain across the entire energy range. The energy and angular resolution together with the ability of the instrument to provide enough information for reliable and efficient discrimination between gamma and hadron-initiated showers shall be substantially improved. In addition, the proposed camera concept shall provide the flexibility and upgradability that all long-lasting experiments seek. The main characteristics of the proposed design are listed in the coming sections. 

\subsection{SiPM based design}

Operation in the presence of high Moon is achieved with PMTs (e.g. for MAGIC and VERITAS), but it needs attention not to damage PMTs and in some cases human intervention to install attenuation filters.
FACT operates for about 30\% longer time than the tagged 'dark time' by MAGIC and even with the Moon shining in the camera without any manual intervention, just increasing the trigger threshold~\cite{FACT_2,Neise:2017ldg}.
Consequently, FACT shows an excellent capability to provide alerts to large telescopes in blazar monitoring even with high Moon.

\begin{figure}[h!]
    \centering
    \includegraphics[width=0.7\textwidth]{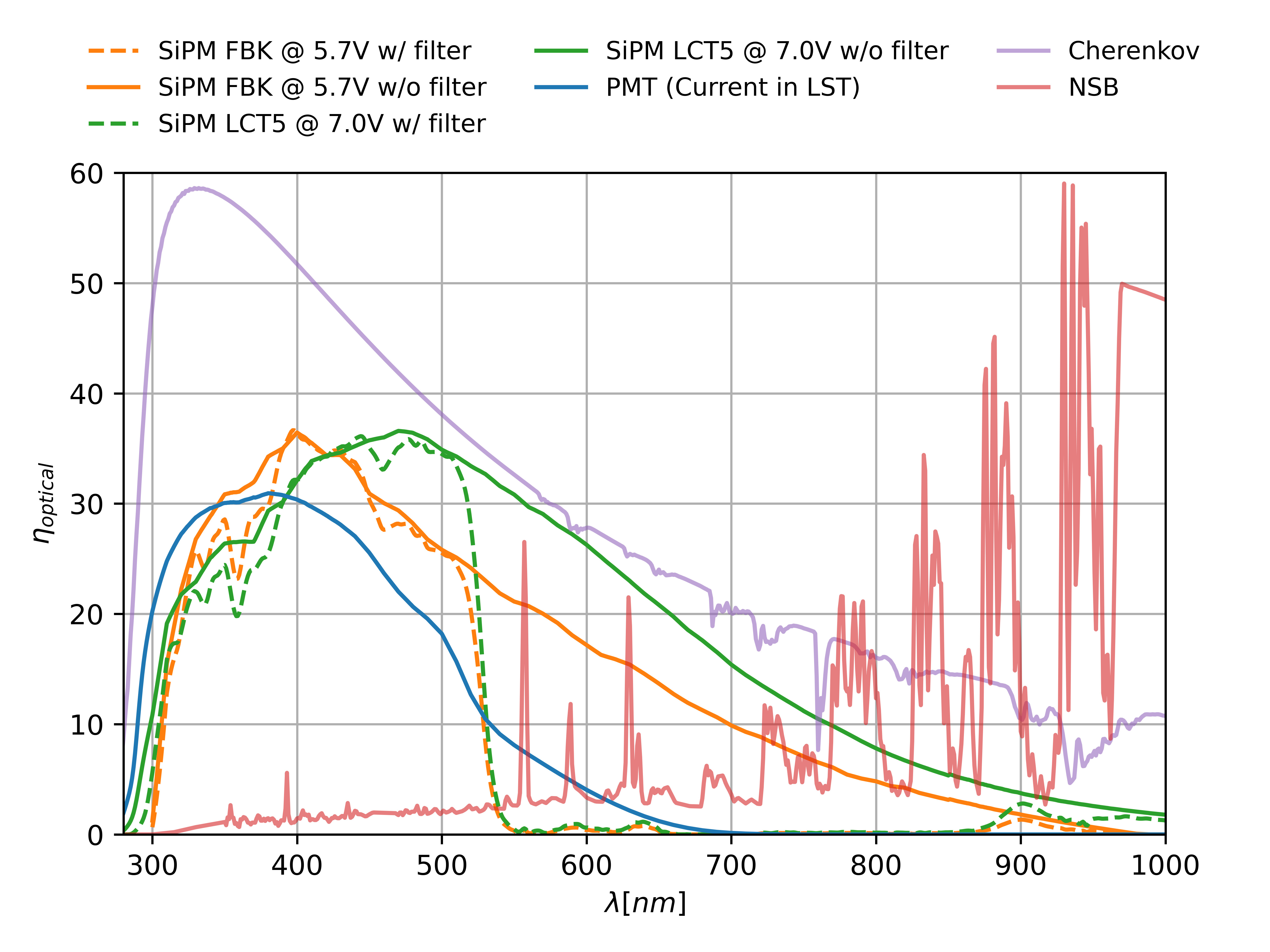}
    \caption{Optical efficiency as function of wavelength for two types of sensors (FBK NUV-HD and Hamamatsu LCT5 technologies) and two entrance windows (standard LST (solid lines), SST-1M filter (dashed lines)). The Cherenkov and reference NSB spectra are shown in arbitrary units as references.}
    \label{fig:optical_efficiency}
\end{figure}

However, replacing PMTs with SiPMs for such a large light collection surfaces is not easy. As a matter of fact, as visible in Figure~\ref{fig:optical_efficiency}, while the quantum efficiency of PMTs drops relatively rapidly above 500~nm, the photodetection efficiency of SiPMs slowly decays until 1~$\mu$m, exhibiting still about 10\% efficiency at 700~nm.
This makes SiPMs more sensitive to background photons. For instance, while the default LST pixel of 0.1$^{\circ}$ equipped with a PMT sees a rate of 0.246~GHz of photo-electrons (p.e.) per pixel in so-called dark conditions at La Palma, the same pixel equipped with an LCT5 SiPM from Hamamatsu operated in optimal conditions (over-voltage of 7~V) would observe 1.4~GHz. 
These background photons will not only affect the resolution of the extracted charge in each pixel but also the hardware threshold at which the telescope will acquire data. 

Several options to reduce the impact of the NSB rate are being studied and will be reported in Sections~\ref{sec:sim_ana} and \ref{sec:pdp}. 

\subsection{Reduced pixel size}
One of the important feature of the project is the use of a smaller pixel size compared to the existing one. 
The IACT sensitivity for point-source searches is strongly influenced by the pixel size, as can be seen using simple cuts on Hillas' parameters of images (see Fig.~5 in Ref.~\cite{funk2009montecarlo}).

The higher granularity of the camera over-samples the image providing more details of the blurred distribution of photons for the more irregular hadronic showers than elongated images of gamma-ray ones. This also applies to muon rings. It also provides more individual time measurements along the shower axis. 
This allows more efficient discrimination between electromagnetic and hadronic extensive air showers, thus improving the sensitivity, as well as energy and angular resolutions.

However the increased number of pixels will affect both the power consumption and data throughput driving naturally the technological choices for the front-end and the digital readout toward integrated circuits.

\subsection{Fully digital readout}

The ambition of this project is to achieve a fully digital readout of all sensors continuously, since this provides 
the desired flexibility and upgradability. 
The acquisition of continuous waveforms enables the use of modern machine learning (ML) techniques in the event selection as close as possible to the sensor. Analysis techniques based on machine learning algorithms have now already proven to be superior to standard ones based on simple cuts on image parameters~\cite{jacquemont2021visapp}.
However, these techniques are normally applied during offline data analysis, while in this project we aim at enabling their use in real-time, for the trigger decision for instance, in order to increase the rejection power and therefore to reduce the energy threshold.
Another approach to reach a fully digital response is to use digital photo-sensors. These sensors would exploit the natural photon counting ability of SiPMs to directly output a photon stream, i.e. a number of photons per time slice of 1~ns. Such sensor tailored for IACT is being designed and will be evaluated during the course of the project. 
The possible paths being investigated to achieve such capability are described in section~\ref{sec:readout}.

\subsection{Design summary}

The key design characteristics of the existing and the proposed camera are listed in Table~\ref{tab:comp_cam}.
\begin{table}[h!]
    \centering
    \begin{tabular}{lr|c|c}
         Parameter & & Current LST camera & Proposed camera  \\
         \hline
         Number of pixels           &               & 1855      & $\sim$7500    \\
         Pixel size                 & [$^{\circ}$]  & 0.1       & 0.05          \\
         \hline
         Number of gain channels             &               & 2         & 1             \\
         Dynamic range              & [p.e.]        & 3000      & $\sim$500           \\
         \hline
         Trigger                    &               & Analogue  & Digital       \\
         Digitization               &               & SCA$^{\star}$+ADC       & FADC          \\
    \end{tabular}
    \caption{Comparison of key characteristics between the existing LST camera and the baseline design for the proposed one based on SiPM sensors.$^{\star}$Switch Capacitor Array. The cameras share the same field of view (4.3$^{\circ}$), sampling speed (1~Giga sample per second (GSPS)) and resolution (12~bit).}
    \label{tab:comp_cam}
\end{table}

\section{Simulation and Analysis for the advanced camera}
\label{sec:sim_ana}

In order to optimize the camera design, an important campaign of simulations is being carried out using CORSIKA and sim\_telarray. 
Among the main parameters to be optimized we can list the following: a) the pixel size and shape; b) the SiPM technology and its operating point; c) the optical properties of the entrance window; d) the pre-amplifying stage performance, i.e. the pulse shape, the gain and signal to noise ratio; and e) the trigger topology. 

Regarding point a), the pixels adopted by the SiPM camera of the SST-1M \cite{Heller2017} have been used as a baseline. They have a hexagonal shape and a linear size 2.5~cm which corresponds to a pixel angular size of 0.05$^\circ$ when used on an LST structure.
Nevertheless, other options for larger pixels will be investigated, as using larger pixels would naturally reduce the channels to instrument in order to obtain the required field of view (FoV). Therefore, this would relax the requirement on the power consumption and data throughput. 
In addition, square pixels would allow to use off-the-shelf SiPM sensors, which saves the additional cost of producing the single photo-mask needed to produce many hexagonal sensors. 
Concerning points b) and c), the optical efficiency as a function of the wavelength of the full photo-detection chain is critical. As seen in Figure~\ref{fig:optical_efficiency}, the sensor choice and the optical properties of the entrance windows have to be optimized together. 
For the time being, two technologies of sensors have been considered: the S13360 (LCT5) technology from Hamamatsu Photonics K. K. (HPK) and the NUV-HD from Fondazione Bruno Kessler (FBK). These are the technologies that have so far shown the best compromise between photo-detection efficiency and speed. 
Two different entrance windows are used in the simulation. The first is the one of the current LST camera and is made of Shinkolite$^\text{TM}$, which is remarkable for its transparency in the UV range. The second entrance window is the one developed for the SST-1M camera, featuring a Borofloat substrate on which an anti-reflective and a dichroic coating are applied. 
The latter is cutting out wavelengths longer than 540~nm. The filter has an impact on the value of the safe threshold~\footnote{The safe threshold is defined as the threshold at which the proton detection rate multiplied by 1.5 and the trigger rate induced by NSB for an NSB rate two times larger than the nominal one intersect.}, as visible in Table~\ref{tab:optical_effi}, namely corresponding to the energy thresholds above which cosmic rays begin to be detectable among the NSB. 
The loss of overall optical efficiency impacts the event reconstruction and affects significantly the telescope performance.

The safe threshold $ST_{pe}$ determined in p.e. cannot be easily compared between the different cameras due to different optical efficiencies. We therefore correct for the optical efficiency of the optical chain (entrance window, light funnels and photo-sensor) to the Cherenkov spectrum $\epsilon_{Cher}$ between 300 and 1000~nm and also account for the artificial boost that the optical cross talk $P_{OCT}$ is giving to the signal:
\begin{equation}
    ST_{\gamma} = \frac{ST_{pe}}{\epsilon_{Cher}\left(1+P_{OCT}\right)}
\end{equation}
The resulting safe thresholds expressed in photons $ST_{\gamma}$ are given in Table~\ref{tab:optical_effi}. They indicate that with a traditional trigger topology based on the signal sum of neighboring pixels, the SiPM performance is comparable to the PMT one or better when operating the SiPM at high values of over-voltage. For instance, working with an optical cross talk of the order of 15\% provides a significant improvement even without the use of a dedicated filter.

\begin{table}[]
    \centering
     \resizebox{\textwidth}{!}{  
    \begin{tabular}{l|c|c|c|c|c|c}
Configuration                   & Safe                & NSB           & Cherenkov         & NSB               & Optical   & Safe      \\
                                &  Threshold	            & rate          & cum. efficiency   & cum. efficiency   & Cross-Talk& Threshold \\
                                & [photons]  & [MHz/pixel]   & [\%]              & [\%]              & [\%]      & [p.e.]      \\
\hline
\hline
PMT w/o filter                  & 286   & 246   & 15.9 & 1.7    & 0    & 45 \\
\hline
LCT5 w/o filter ($\Delta$V=4.4) & 272   & 386   & 20.8 & 6.2   & 8     & 61 \\ 
LCT5 w/ filter  ($\Delta$V=4.4) & 229   & 108   & 14.0 & 1.9   & 8     & 34 \\ 
\hline
LCT5 w/o filter ($\Delta$V=7.0) & 218   & 426   & 24.3 & 8.1   & 15    & 67 \\ 
LCT5 w/ filter  ($\Delta$V=7.0) & 179   & 109   & 16.0 & 2.5   & 15    & 36 \\ 
    \end{tabular}
    }
    \caption{Summary of safe threshold calculation. These thresholds are achieved for 3~ns FWHM pulses. The Cherenkov and NSB efficiencies are obtained by normalizing the area under their spectrum over the wavelength range from 300 to 1000~nm and multiplying by the optical efficiency of a single pixel. The equivalent FoV of the trigger patches are 0.46$^\circ$ for the standard camera (21 neighboring pixels) and 0.35$^\circ$ for the proposed one (49 neighboring pixels)}
    \label{tab:optical_effi}
\end{table}

The point d) is addressed by varying the signal to noise ratio and the pulse width. As foreseen, with the default trigger strategy, achieving a shorter pulse limits the consequences of the pile-up and allows to significantly reduce the NSB-induced trigger rate.
However, studies for point e) might reveal that novel strategies for trigger may reduce the importance of having a narrow pulse and relax the requirement on the sensor choice and pre-amplifying stage speed.

The use of ML algorithms, which are very effective for image or time series analysis, would better exploit all image features contrary to the standard Hillas parametrization. For this reason, two analysis pipelines~\cite{ctlearn, jacquemontCBMI_arxiv} based on convolutional neural networks are being tested for the image reconstruction. In addition other methods are been investigated, like look-up tables or semi-analytical formula~\cite{cat_telescope_reco, impact_analysis, lhfit, de_Naurois_2009}.
These methods have been yet only developed for single-telescope observations, while the performance gain that these camera brings is probably best for stereoscopic analysis of multiple telescope images. Porting these algorithms to stereoscopic analysis is also part of the work on-going in this project.

\subsection{Photon-detection plane}
\label{sec:pdp}

As described in the previous section, using a fast and low-power pre-amplifying stage is critical for this project. 
A pre-amplifying stage has been developed by INFN based on bipolar junction transistors~\cite{infn_preamp} and used for measurements with 0.1$^\circ$ and 0.05$^\circ$ pixels. It achieves remarkably short pulses as seen in Tab.~\ref{tab:pulse_shape}. While this stage meets the requirements on the speed, it would require about 3.2~kW for 7500 pixels that equipped the SiPM camera. This would leave no margin on the power to accommodate the remaining components of the readout chain such as FADCs, FPGAs and optical transceivers. On the other hand, the MUSIC ASIC\cite{MUSIC} would meet the power consumption requirements but it exhibits pulse widths yet too large to limit the NSB pile-up effect.

\begin{table}[h!]
    \centering
     \resizebox{\textwidth}{!}{  
    \begin{tabular}{l|c|c|c}
Sensor type &	pre-amplifying stage & FWHM & Power consumption \\
       &                        & [ns] & [mW/pixel] \\
\hline
FBK NUV-HD (14$\times$ 6$\times$6~mm$^2$)   &  INFN     &   3.4   & 1500 \\
FBK NUV-HD (12$\times$ 6$\times$6~mm$^2$)   &  MUSIC    &   5.3   & 830 \\ 
HPK LCT2 (hex. sensor)                      &  INFN     &   2.9   & 450 \\ 
HPK LCT2 (hex. sensor)                      &  MUSIC    &   5.1   & 100 \\ 
    \end{tabular}
    }
    \caption{Summary of pre-amplifying stage performance. The power consumption here is an estimate and does not include power dissipated by the DC/DC converters required.}
    \label{tab:pulse_shape}
\end{table}

Therefore, this project targets the development of an ASIC.
Its design will be based on recent studies~\cite{Delacour} and state of the art commercial input stages from the MUSIC ASIC and its successor, the FastIC. 
This ASIC will be tailored for the SiPM technology chosen, even though it will incorporate variable parameters in order to accommodate for other sensor technologies which matters for possible upgrades.

\subsection{Camera readout}
\label{sec:readout}

The readout architecture of the camera will play a major role to achieve the expected performance boost and allow for possible upgrades during the detector lifetime. 

High speed FADCs coupled to high performance FPGAs have been widely used in particle physics and astrophysics. For instance, the SST-1M camera features a fully digital readout system~\cite{digicam} in which the signals from SiPMs are sent to FADCs, converted and streamed to an FPGA where a digital algorithm is used to select data to send to the disk.

The camera proposed here is an evolution of this concept, whereas the FADC sampling frequency is much higher, at least 1~GSPS, the number of channels is a factor of four higher and the rate (and therefore the throughput) is one order of magnitude higher.

Commercial FADC operating at GSPS frequencies with a serial output exist, but their number of channels is limited, resulting in high power consumption and high cost per channel.
Therefore, a custom solution with more channels per chip and with a power consumption of less than 0.1~W/channel needs to be developed.  
ASICs can achieve this goal with heavy use of event-driven operation and ultra-low-power digital circuits implemented in CMOS technologies that enable miniaturization, speed, and low power, simultaneously. 
Several solutions are considered and the corresponding architectures are being analyzed: 

\paragraph{\bf Trigger-less readout}

The most appealing readout which also follows the trend of many detectors in high-energy physics, is the so-called trigger-less.
Continuous digitization and shipping out of the signals of each camera allows to form triggers of more telescopes before a single trigger decision is taken. This will increase substantially the performance of the advanced SiPM cameras with respect to the current scheme of the PMT camera.

The disadvantage of such solution is that the amount of data to be transferred from the camera to the camera server is tremendous. Assuming 7500 pixels sampled at 1~GSPS with 12 bits resolution, data should be transferred at $\sim$90~Tb/s. Two solutions are being explored in order to cope with such high throughput. The first one focuses on high-end technologies for optical transfer and presents the advantage that most of the complexity and power consumption is transferred away from the camera to the camera server. The second one consists of implementing on-the-fly data volume reduction using ML algorithms (e.g. auto-encoders, inverse neural networks), which requires high processing power inside the camera.
This solution is the preferred one as it offers potentially the highest performance gain, flexibility and upgradability.

\paragraph{\bf Advanced trigger readout}
Two alternative solutions to reduce the required data throughput are also being considered.
The first one relies on the use of high performance FPGAs to implement advanced trigger algorithms, which offer higher background rejection. 
Several trigger algorithms are being studied, either based on digital filters or machine learning algorithms and all aim at analyzing data cubes instead of images.
Always with the aim of reducing the data throughput, the second solution envisaged would use a two level trigger: one low-level implemented in the camera which would ensure the highest acceptable throughput and a high-level one running in the camera server for more refined trigger decision. This second approach has the advantage of transferring the complexity from the camera to the camera server and therefore limits the power dissipation inside the camera.

\paragraph{\bf Digital sensor}

The idea of having a sensor which directly outputs a photon stream is very appealing, as it would simplify the architecture, relax most of the requirements in term of data throughput and power consumption. 
A Digital Photon Counter (DiPC) tailored for the use in IACT cameras is being developed. The basic idea is to identify the leading edge of a discharging micro-cells and add the numbers with a clock frequency of about 1~GHz. The sum will be stored in an on-chip memory, while reduced precision information will be sent out to the digital camera trigger. 
In case of a positive trigger decision, the high precision data can be read asynchronously with limited bandwidth. 
The very precise arrival time information for each individual photons will allow the reconstruction of the three-dimensional information from each photon, converting the differences in arrival time to a distance along the trajectory and most probably greatly improving background rejection and angular resolution.

\section{Conclusions}

The design of a novel camera for the LST is on-going and challenges on multiple fronts are being addressed by an international collaboration. The important simulation effort deployed will soon allow to select the suited entrance window and light guides, the right sensor and its optimal working point as well as the pixel size and geometry. Combined with studies of novel trigger approaches and analysis pipelines, the technical specifications for preamplification stage and the digital readout will be clarified. The design and production of tailored ASICs for the preamplification and the digitizing stages will naturally follow. Eventually, the stereoscopic data analysis pipelines will shed light on the ultimate performance, which such innovative camera will bring to the field of IACT.

\section{Acknowledgment}
We gratefully acknowledge financial support from the agencies and organizations listed here: www.cta-observatory.org/consortium\_acknowledgments

%%%%%%%%%%%%%%%   BIBLIOGRAPHY    %%%%%%%%%%%%%%%%%%%%%%

\bibliographystyle{mnras}
\bibliography{main} 

\begin{thebibliography}{}
\makeatletter
\relax
\def\mn@urlcharsother{\let\do\@makeother \do\$\do\&\do\#\do\^\do\_\do\%\do\~}
\def\mn@doi{\begingroup\mn@urlcharsother \@ifnextchar [ {\mn@doi@}
  {\mn@doi@[]}}
\def\mn@doi@[#1]#2{\def\@tempa{#1}\ifx\@tempa\@empty \href
  {http://dx.doi.org/#2} {doi:#2}\else \href {http://dx.doi.org/#2} {#1}\fi
  \endgroup}
\def\mn@eprint#1#2{\mn@eprint@#1:#2::\@nil}
\def\mn@eprint@arXiv#1{\href {http://arxiv.org/abs/#1} {{\tt arXiv:#1}}}
\def\mn@eprint@dblp#1{\href {http://dblp.uni-trier.de/rec/bibtex/#1.xml}
  {dblp:#1}}
\def\mn@eprint@#1:#2:#3:#4\@nil{\def\@tempa {#1}\def\@tempb {#2}\def\@tempc
  {#3}\ifx \@tempc \@empty \let \@tempc \@tempb \let \@tempb \@tempa \fi \ifx
  \@tempb \@empty \def\@tempb {arXiv}\fi \@ifundefined
  {mn@eprint@\@tempb}{\@tempb:\@tempc}{\expandafter \expandafter \csname
  mn@eprint@\@tempb\endcsname \expandafter{\@tempc}}}

\bibitem[\protect\citeauthoryear{Berti et~al.,}{Berti
  et~al.}{2020}]{infn_preamp}
Berti A.,  et~al., 2020, \mn@doi [NIMA]
  {https://doi.org/10.1016/j.nima.2020.164373}, 982, 164373

\bibitem[\protect\citeauthoryear{{Biland}}{{Biland}}{2014}]{FACT_2}
{Biland} A. e. a. F.~C.,  2014, \mn@doi [JINST]
  {10.1088/1748-0221/9/10/p10012}, 9 P10012

\bibitem[\protect\citeauthoryear{Delacour}{Delacour}{2020}]{Delacour}
Delacour C.,  {2020}, {\href{https://cds.cern.ch/record/2744108 }{Design of an
  analogue front-end for Silicon Photo Multipliers}}

\bibitem[\protect\citeauthoryear{{Emery} et~al.}{{Emery} et~al.}{2021}]{lhfit}
{Emery} G.,  et~al., 2021, in in these proceedings, https://pos.sissa.it/395.

\bibitem[\protect\citeauthoryear{Funk \& Hinton}{Funk \&
  Hinton}{2009}]{funk2009montecarlo}
Funk S.,  Hinton J.,  2009, \mn@doi [AIP Conf. Proc.] {10.1063/1.3076817},
  1085, 878

\bibitem[\protect\citeauthoryear{G{\'o}mez et~al.}{G{\'o}mez
  et~al.}{2016}]{MUSIC}
G{\'o}mez S.,  et~al., 2016, in Photonics Europe.

\bibitem[\protect\citeauthoryear{Heller et~al.,}{Heller
  et~al.}{2017}]{Heller2017}
Heller M.,  et~al., 2017, \mn@doi [The European Physical Journal C]
  {10.1140/epjc/s10052-017-4609-z}, 77, 47

\bibitem[\protect\citeauthoryear{Jacquemont et~al.}{Jacquemont
  et~al.}{2021a}]{jacquemont2021visapp}
Jacquemont M.,  et~al., 2021a, in Proceedings of VISIGRAPP. SciTePress

\bibitem[\protect\citeauthoryear{{Jacquemont} et~al.}{{Jacquemont}
  et~al.}{2021b}]{jacquemontCBMI_arxiv}
{Jacquemont} M.,  et~al., 2021b, arXiv e-prints, \href
  {https://ui.adsabs.harvard.edu/abs/2021arXiv210514927J} {p. arXiv:2105.14927}

\bibitem[\protect\citeauthoryear{Le~Bohec et~al.,}{Le~Bohec
  et~al.}{1998}]{cat_telescope_reco}
Le~Bohec S.,  et~al., 1998, \mn@doi [NIMA] {10.1016/s0168-9002(98)00750-5},
  416, 425–437

\bibitem[\protect\citeauthoryear{{Mazin} et~al.}{{Mazin}
  et~al.}{2021}]{lst_mazin}
{Mazin} D.,  et~al., 2021, in in these proceedings, https://pos.sissa.it/395.

\bibitem[\protect\citeauthoryear{Neise et~al.}{Neise
  et~al.}{2017}]{Neise:2017ldg}
Neise D.,  et~al., 2017, \mn@doi [Nucl. Instrum. Meth.]
  {10.1016/j.nima.2016.12.053}, A876, 17

\bibitem[\protect\citeauthoryear{Nieto et~al.}{Nieto et~al.}{2021}]{ctlearn}
Nieto D.,  et~al., 2021, Reconstruction of IACT events using deep learning
  techniques with CTLearn (\mn@eprint {arXiv} {2101.07626})

\bibitem[\protect\citeauthoryear{Parsons \& Hinton}{Parsons \&
  Hinton}{2014}]{impact_analysis}
Parsons R.,  Hinton J.,  2014, \mn@doi [Astroparticle Physics]
  {10.1016/j.astropartphys.2014.03.002}, 56, 26–34

\bibitem[\protect\citeauthoryear{Rajda et~al.}{Rajda et~al.}{2016}]{digicam}
Rajda P.,  et~al., 2016, \mn@doi [PoS] {10.22323/1.236.0931}, ICRC2015, 931

\bibitem[\protect\citeauthoryear{de Naurois \& Rolland}{de~Naurois \&
  Rolland}{2009}]{de_Naurois_2009}
de Naurois M.,  Rolland L.,  2009, \mn@doi [Astroparticle Physics]
  {10.1016/j.astropartphys.2009.09.001}, 32, 231–252

\makeatother
\end{thebibliography}

%% Full authors list (ONLY FOR COLLABORATIONS)
\clearpage
\section*{Full Authors List: \Coll\ Collaboration}

\scriptsize
\noindent
H. Abe$^{1}$,
A. Aguasca$^{2}$,
I. Agudo$^{3}$,
L. A. Antonelli$^{4}$,
C. Aramo$^{5}$,
T.  Armstrong$^{6}$,
M.  Artero$^{7}$,
K. Asano$^{1}$,
H. Ashkar$^{8}$,
P. Aubert$^{9}$
% ...
\\

\noindent
$^{1}$Institute for Cosmic Ray Research, University of Tokyo.
$^{2}$Departament de Física Quàntica i Astrofísica, Institut de Ciències del Cosmos, Universitat de Barcelona, IEEC-UB.
$^{3}$Instituto de Astrofísica de Andalucía-CSIC.
$^{4}$INAF - Osservatorio Astronomico di Roma.
$^{5}$INFN Sezione di Napoli.
$^{6}$Aix Marseille Univ, CNRS/IN2P3, CPPM.
$^{7}$Institut de Fisica d'Altes Energies (IFAE), The Barcelona Institute of Science and Technology.
$^{8}$IRFU, CEA, Université Paris-Saclay.
$^{9}$LAPP, Univ. Grenoble Alpes, Univ. Savoie Mont Blanc, CNRS-IN2P3, Annecy.
% ...
%
%\noindent \textbf{Note comment afterwards:} Collaborations have the possibility to provide an authors list in xml format which will be used while generating the DOI entries making the full authors list searchable in databases like Inspire HEP. For instructions please go to icrc2021.desy.de/proceedings or contact us under icrc2021proc@desy.de.\\
%
%\scriptsize
%\noindent
%first.author$^1$, 
%second.author$^2$, 
%third.author$^3$ % .... more names
%and 
%last.author$^{n}$ \\
%
%\noindent
%$^1$first.affiliation.
%$^2$second.affiliation. % .... more affiliation
%$^{m}$last.affiliation.

\end{document}